\documentclass[]{aa}
\usepackage{graphicx}
\usepackage{psfig}
\usepackage{epstopdf}
\usepackage{amssymb}
\usepackage{amsmath}
\usepackage{verbatim}
\usepackage{natbib}
\usepackage{rotate}
\usepackage{lscape}
\usepackage{aalongtable}
\usepackage{supertabular}
\usepackage{mathbbol}
\usepackage{bm}
\usepackage{mathtools}
\usepackage{enumerate}
\usepackage{arydshln}
\usepackage{amsmath}

\bibpunct{(}{)}{;}{a}{}{,}

\newcommand{\appropto}{\mathrel{\vcenter{
  \offinterlineskip\halign{\hfil$##$\cr
    \propto\cr\noalign{\kern2pt}\sim\cr\noalign{\kern-2pt}}}}}
\newcommand{\FigDir}{}

\begin{document}      

\title{\vspace{-5cm}\\
Applying Wirtinger derivatives to the radio interferometry calibration problem}

\subtitle{}
\author{C. Tasse\inst{1,2}}

\institute{
GEPI, Observatoire de Paris, CNRS, Universit\'e Paris Diderot,
5 place Jules Janssen, 92190 Meudon, France
\and
Department of Physics \& Electronics, Rhodes University, PO Box 94,
Grahamstown, 6140, South Africa
}

\abstract{This paper presents a fast algorithm for full-polarisation,
  direction dependent calibration in radio interferometry. It is based
  on Wirtinger's approach to complex differentiation. Compared to the
  classical case, and under reasonable approximations, the Jacobian
  appearing in the Levenberg-Maquardt iterative scheme presents
  a sparse structure, allowing for high gain in terms of algorithmic cost.
}

\authorrunning{C. Tasse}

\titlerunning{Applying Wirtinger derivatives to the radio interferometry calibration problem}
   \maketitle

\def\COH{{\sc CohJones}}

\def\u{u}
\def\v{v}
\def\w{w}
\def\l{l}
\def\m{m}
\def\n{n}
\def\d{d}
\def\dbf{{\bm d}}

\newcommand{\conj}[1]{\overline{#1}}
\newcommand{\conjp}[1]{\left({\overline{#1}}\right)}

\def\JMat{\textbf{J}}
\def\Skyd{\textbf{S}_d}
\def\Vbl{\textbf{V}_{(pq)t\nu}}

\newcommand{\mat}[1]{{\bm{#1}}}
\newcommand{\JJ}{\mat{J}} 
\newcommand{\JHJ}{\JJ^H\JJ} 

\def\vbltnu{\textbf{v}_{(pq)t\nu}}
\def\vbl{\textbf{v}_{pq}}
\def\V{\textbf{V}}
\def\gwirt{\bm{g_{\!_{W}}}}
\def\dgwirt{\dbf\bm{g_{\!_{W}}}}
\def\g{\bm{g}}
\def\vis{\textbf{v}}

\def\separator{
\hrule
\begin{center}
\textsc{to be modified after that}
\end{center}
\hrule
}

\def\Kron{\otimes}

\def\SimpleJacob{\bm{J}}
\def\Jacob{\bm{\mathcal{J}}}
\def\JVpq{\Jacob_{{\textbf{v}_{pq}},\bm{g_{\!_{W}}}}}

\def\JVpqg{\Jacob_{{\textbf{v}_{pq}},\bm{g}}}
\def\JVpqCg{\Jacob_{{\textbf{v}_{pq}},\bm{\conj{g}}}}
\def\JVAtg{\JV\big|_{\vec{g}}}

\def\A{\textbf{A}}
\def\H{\textbf{H}}

\def\JV{\Jacob\left\{\textbf{v}\right\}}
\def\JV{\Jacob_{\textbf{v}}}
\def\JVg{\Jacob_{\textbf{v},\bm{g}}}
\def\JVCg{\Jacob_{\textbf{v},\bm{\conj{g}}}}

\section{Complex optimisation}
\label{sec:Wirtinger}

Given a set of cross correlations between antenna voltages (the
visibility set) in a given time-frequency domain and a model of the sky brightness, the calibration
step in radio interferometry consists in estimating a set of
direction-dependent Jones matrices per antenna. This is usually done
using a chi-square minimisation technique such as the
Levenberg-Maquardt. In this paper, I use the alternative Wirtinger's
definition of complex derivative, instead of what is usually done by
considering the real and imaginary (later referred as the
{\it Wirtinger} and {\it classical} approaches respectively).

\subsection{RIME formalism}

Using the Radio Interferometry Measurement Equation (RIME) formalism
\citep[][]{Hamaker96}, the
4-polarisation visibility vector $\vbl$ measured on baseline $(pq)$,
at time $t$ and frequency $\nu$ can be written as

\begin{alignat}{2}
\label{eq:ME}
\vbl=&\text{Vec}\left(\Vbl\right)&\\
=&\displaystyle\sum\limits_{d} \left(\conj{\JMat^{d}_{qt\nu}}\Kron \JMat^{d}_{pt\nu}\right)
\text{Vec}\left(\Skyd\right) k^{d}_{(pq)t\nu}\\
\text{with}\ k^{d}_{(pq)t\nu}=&\exp{\left(-2 i\pi
  \left(\u\l+\v\m+\w\left(\text{n}-1\right)\right)\right)}\\
\text{and}\ \n=&\sqrt{1-\l^2-\m^2}
\end{alignat}

\noindent where $\JMat^{d}_{pt\nu}$ is the Jones matrix of antenna $p$
in direction $d$, $\Skyd$ is the four-polarization sky matrix, $\Kron$
is the Kronecker product, $[\u,\v,\w]^T$ is the baseline vector between antennas
$p$ and $q$ in wavelength units, and
$\vec{s}_d=[\l,\m,\n=\sqrt{1-\l^2-\m^2}]^T$ is a sky direction later
labeled as $d$.
In the following, the Jones matrix $\JMat^{d}_{pt\nu}$ is
represented by scalars $g^{d}_{pt\nu,k}$ as

\begin{equation}
\JMat^{d}_{pt\nu}=
\begin{bmatrix}
g^{d}_{pt\nu,0} & g^{d}_{pt\nu,2} \\ 
g^{d}_{pt\nu,1} & g^{d}_{pt\nu,3} 
\end{bmatrix}
\end{equation}

\noindent and the set of Jones matrices is represented by a gain
vector $\vec{g}$ containing the scalars $g^{d}_{pt\nu,k}$ for all
directions, antennas and polarisations.

From Eq. \ref{eq:ME}, the $i^{th}$ polarisation component of $\vbl$
can be written as

\begin{alignat}{2}
\label{eq:hpq}
\mathrm{v}^i_{pqt\nu}=&h^i_{pq}(\vec{g})\\
=&
\displaystyle\sum\limits_{d}
\displaystyle\sum\limits_{j=0}^3 \left(g^{d}_{pt\nu,\textbf{A}_{ij}}.\conj{g^{d}_{qt\nu,\textbf{B}_{ij}}}\right).k^{d}_{pqt\nu}.\text{s}_{d,j}
\end{alignat}

\noindent where 

\begin{equation}
\textbf{A}=
\begin{bmatrix}
0 & 2 & 0 & 2 \\ 
1 & 3 & 1 & 3 \\ 
0 & 2 & 0 & 2 \\
1 & 3 & 1 & 3 
\end{bmatrix}
\text{ and }
\textbf{B}=
\begin{bmatrix}
0 & 0 & 2 & 2 \\ 
0 & 0 & 2 & 2 \\ 
1 & 1 & 3 & 3 \\
1 & 1 & 3 & 3 
\end{bmatrix}
\end{equation}

\subsection{Wirtinger complex derivative}
\label{sec:Cderiv}

In order to compute a Jacobian, a
derivative definition for complex numbers has to be chosen. Instead of
differentiating against real and imaginary parts independently, one can
adopt a Wirtinger differentiation point of view and consider the
complex and their conjugate as being independent. Choosing this type of differentiation
turns out to be rather powerful to solve problems of the form of
Eq. \ref{eq:ME} (see
Sec. \ref{sec:Solver}). If a complex number is written as $z=x+iy$, the
Wirtinger complex derivative operator becomes

\begin{alignat}{3}
\frac{\partial }{\partial z}&=&\frac{1}{2}\left(\frac{\partial }{\partial x}-i\frac{\partial }{\partial y}\right)\\
\text{and }\frac{\partial }{\partial \conj{z}}&=&\frac{1}{2}\left(\frac{\partial }{\partial x}+i\frac{\partial }{\partial y}\right)
\end{alignat}

\noindent where $x$ and $y$ are the real and imaginary parts
respectively. The
Wirtinger has a trivial but remarkable property that a scalar and its
complex conjugate can be viewed as independent variables, and in
particular


\begin{equation}
\label{eq:propConj}
\frac{\partial \conj{z}}{\partial z}=0
\text{ and }
\frac{\partial z}{\partial \conj{z}}=0
\end{equation}

Considering the sky, gain, and geometry relation given in
Eq. \ref{eq:ME}, according to the property of Wirtinger derivative of complex
conjugate (Eq. \ref{eq:propConj})

\begin{alignat}{3}
\label{eq:Deriv_gp}
\frac{\partial \mathrm{v}^i_{pqt\nu}}{\partial g^{d}_{pt\nu,\textbf{A}_{ij}}}=&
\left(\conj{g^{d}_{qt\nu,\textbf{B}_{ij}}}\text{s}_{d,j}\right).k^{d}_{pqt\nu}\\
\label{eq:Deriv_gq}
\text{and}\ 
\frac{\partial \mathrm{v}^i_{pqt\nu}}{\partial g^{d}_{qt\nu,\textbf{B}_{ij}}}=&0
\end{alignat}

\noindent while differentiating against the complex conjugate of those
variables, one obtain

\begin{alignat}{3}
\label{eq:Deriv_Cgp}
\frac{\partial \mathrm{v}^i_{pqt\nu}}{\partial \conjp{g^{d}_{pt\nu,\textbf{A}_{ij}}}}=&0\\
\label{eq:Deriv_Cgq}
\text{and}\ 
\frac{\partial \mathrm{v}^i_{pqt\nu}}{\partial \conjp{g^{d}_{qt\nu,\textbf{B}_{ij}}}}=&
\left(g^{d}_{pt\nu,\textbf{A}_{ij}}\text{s}_{d,j}\right).k^{d}_{pqt\nu}
\end{alignat}

Interestingly, Eq. \ref{eq:Deriv_gp},
\ref{eq:Deriv_gq}, \ref{eq:Deriv_Cgp} and \ref{eq:Deriv_Cgq} show that the
derivatives are always constant with respect to the differential
variable.

\subsection{Wirtinger Jacobian}
\label{sec:WirtingerJacob}

This section describes the structure of the Wirtinger Jacobian $\JV$ using the
results of Sec. \ref{sec:Cderiv}.
First, let consider the visibility vector $\vbl$
for all given time frequency blocks, and write the antenna,
polarisation, and direction dependent gain vector as
$\vec{g}$. Its size is $4n_an_d$, and have for $k^{th}$ component
$k=j+4\times d+4\times a \times n_d$ the gain of antenna $a$ in
direction $d$ for polarisation $j$, where $n_a$, and $n_d$ are the
number of antenna and directions.

The corresponding Wirtinger Jacobian $\JVpq$ has size $(4n_t n_{\nu})\times (8n_a n_d)$ ($n_t$
and $n_{\nu}$ are the number of time and frequency points), and can be
decomposed as follows:

\begin{alignat}{2}
\label{eq:dV}
\dbf\vbl=&\JVpq\ \dbf\gwirt\\
=&\JVpqg\ \dbf\g+\JVpqCg\ \dbf\conj{\g}\\
\text{where }\dgwirt=&
\begin{bmatrix}
\dbf\vec{g}\\
\dbf\conj{\vec{g}}
\end{bmatrix}\\
\text{and }\JVpq=&
\begin{bmatrix}
\JVpqg & \JVpqCg
\end{bmatrix}
\end{alignat}




Each cell of
$\JVpq$, $\JVpqg$ and $\JVpqCg$ can be written using Eq. \ref{eq:Deriv_gp}, \ref{eq:Deriv_gq},
\ref{eq:Deriv_Cgp} and \ref{eq:Deriv_Cgq}. Specifically, line corresponds to a
single $i$-polarisation measurement at $(t\nu)$ for the $(pq)$ baseline, and a column
$j+4d+a.4n_d$ to a gain for polarisation $j$, antenna $a$ and
direction $d$ ($g^{d}_{a,j}$). The matrix $\JVpq$ can be described as


\begin{alignat}{2}
\label{eq:Jpq0}
\left[\JVpqg\right]_{t\nu,i}=&
\begin{cases}
\left(\conj{g^{d}_{qt\nu,\textbf{B}_{ij}}}\text{s}_{d,j}\right).k^{d}_{pqt\nu}\text{ for }a=p\\
0\ \text{otherwise}
\end{cases}
\end{alignat}

\noindent and

\begin{alignat}{2}
\label{eq:Jpq1}
\left[\JVpqCg\right]_{t\nu,i}=&
\begin{cases}
\left(g^{d}_{pt\nu,\textbf{A}_{ij}}\text{s}_{d,j}\right).k^{d}_{pqt\nu}\text{ for }a=q\\
0\ \text{otherwise}
\end{cases}
\end{alignat}

\noindent One can see that non-zero columns are the ones
corresponding to all direction and all polarisations for antenna $p$. The Jacobian for all
baselines is written in a similar way, by superposing the
$\JVpq$ for all $(pq)$ pairs as follows:

\begin{alignat}{2}
\label{eq:J}
\JV=&
\begin{bmatrix} 
\vdots \\ 
\JVpq\\ 
\vdots \\ 
\end{bmatrix}
\end{alignat}

\noindent which have size $[(4n_{bl} n_t n_{\nu})\times (8n_a n_d)]$,
where $n_{bl}$ is the number of baselines and is typically
$n_{bl}=n_a(n_a-1)/2$. Although it has large dimensions, $\JV$ is
sparse.


\section{Fast iterative solver using Wirtinger's framework}
\label{sec:Solver}

\begin{figure*}[t!]
\begin{center}
\includegraphics[width=\textwidth]{\FigDir 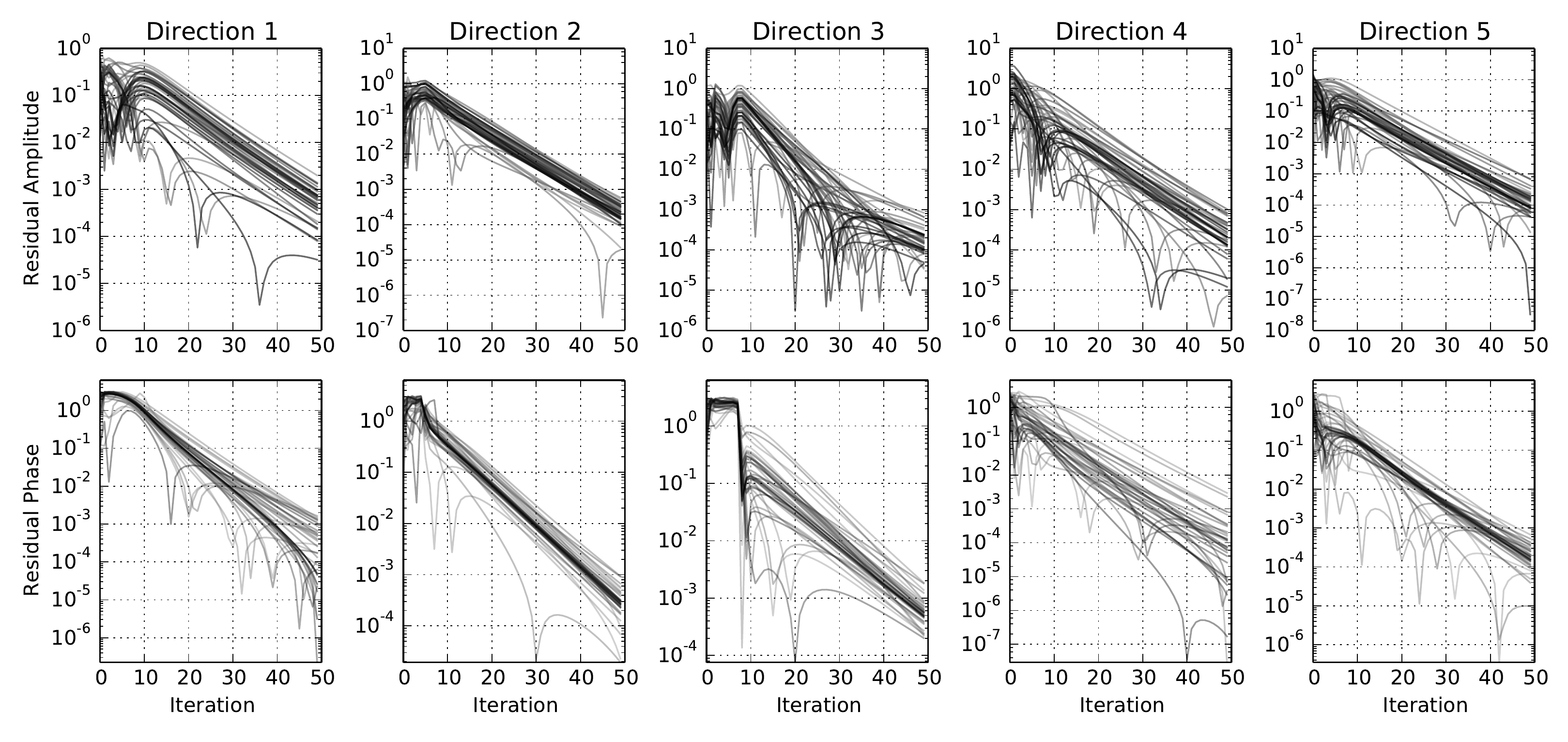}
\caption{\label{fig:Convergence} This plot shows the amplitude (top
  panels) and
  phase (bottom panels) of the difference between
  the estimated gains and the true (random) gains in the different
  directions for the different antenna (shaded greys full lines).}
\end{center}
\end{figure*}

This section describes a Levenberg-Maquardt based calibration algorithm, using
Wirtinger derivative to define the Jacobian (the \COH~algorithm, see
Sec. \ref{sec:COH}). This problem is addressed in greater detail in \citet[][]{SmirnovTasse14}.

\subsection{Levenberg-Maquardt}

In the following $h$ is the non-linear operator mapping the
4-polarisations, direction-dependent gain vector $\vec{g}$ to the
visibility vector $\vis$ containing all baselines, time and frequency
data

\begin{alignat}{2}
\vis=h\left(\vec{g}\right)=&
\begin{bmatrix} 
\vdots \\ 
h_{pq}\left(\vec{g}\right)\\ 
\vdots \\ 
\end{bmatrix}
\end{alignat}

\noindent where $h_{pq}$ is defined in Eq. \ref{eq:hpq}.

The direction-dependent Jones matrices appearing in the measurement
equation (Eq. \ref{eq:ME}) can then be estimated using a chi-square
minimisation technique such as the Levenberg-Maquardt. Specifically, the gain vector $\vec{g}$ can be iteratively estimated from the
measured visibilities $\mathbf{v_m}$ as

\def\SimpleJacobAtXi{\bm{J}\big|_{\widehat{\g_{i}}}}
\def\HAtXi{\H\big|_{\widehat{\g_{i}}}}
\def\KAtXi{\textbf{K}\big|_{\widehat{\g_{i}}}}
\begin{alignat}{2}
\label{eq:LM}
\widehat{\vec{g}_{i+1}}=&\widehat{\vec{g}_{i}}+\KAtXi^{-1}\SimpleJacobAtXi^H\textbf{C}^{-1}\left(\mathbf{v_m}-h\left(\widehat{\vec{g}_{i}}\right)\right)\\
\text{with }\KAtXi=&\HAtXi+\lambda.\text{diag} \left(\HAtXi\right)\\
\text{and }\HAtXi=&\SimpleJacobAtXi^H\textbf{C}^{-1}\SimpleJacobAtXi
\end{alignat}

\noindent where the matrix $\SimpleJacobAtXi$ is the Jacobian of $h$
estimated at $\widehat{\g_{i}}$, and
$\textbf{C}$ is the covariance matrix of $\mathbf{v}$.

The Levenberg-Maquardt algorithm can be equivalently applied by using the Wirtinger Jacobian
or the classical Jacobian. In this paper, as mentioned in Sec. \ref{sec:Wirtinger}, instead of computing the Jacobian using the real and
imaginary parts of gains as independent variables, we use the Wirtinger
derivative definition (see Sec. \ref{sec:WirtingerJacob})

\begin{alignat}{2}
\label{eq:LMWirtinger}
\SimpleJacob:=&\JV
\end{alignat}

\subsection{The \COH~algorithm}
\label{sec:COH}

\begin{figure}[h!]
\begin{center}
\includegraphics[width=\columnwidth]{\FigDir 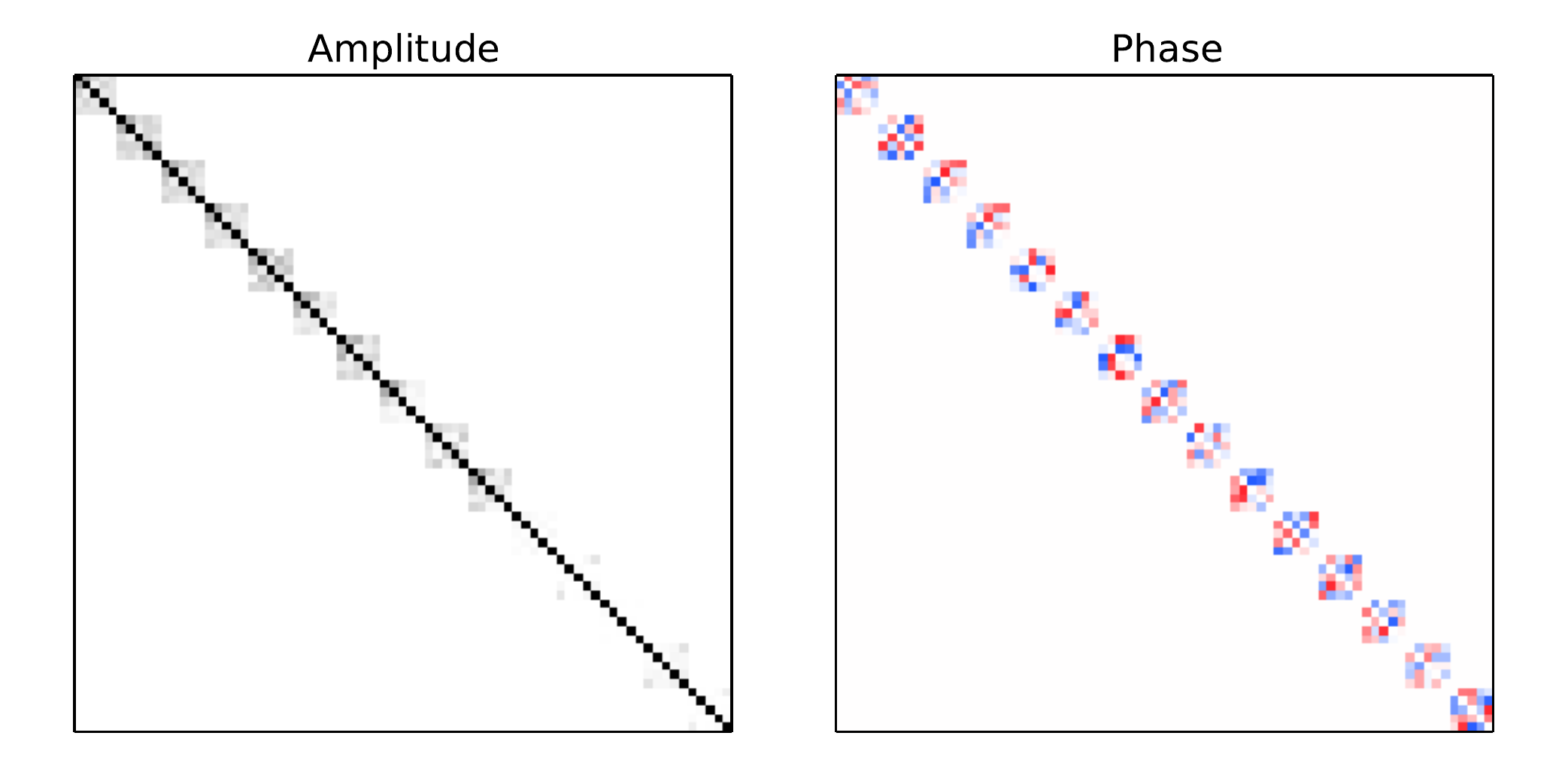}
\caption{\label{fig:HalfJHJ} This figure shows amplitude (left panel)
  and phase (right panel) of the block-diagonal matrix $\JVg^H\JVg$
  for the dataset described in the text. Each block corresponds to the
different directions for each specific antenna. Its block structure
make it easily invertible.}
\end{center}
\end{figure}

The Wirtinger $\SimpleJacob^H\SimpleJacob$ has a different structure to the classical one. In this section, we describe an algorithms that uses {\it only one} of the
two independent Wirtinger variable (either $z$ or $\conj{z}$). In
short, in this section $\SimpleJacob$ is reduced to

\begin{alignat}{2}
\label{eq:LMWirtinger}
\SimpleJacob:=&\JVg
\end{alignat}

\noindent which means only half of the
Jacobian $\JV$ is used, and a single principal block in the matrix
$\JV^H\JV$ is constructed (see Fig. \ref{fig:HalfJHJ}). The algorithm is referred to {\it Complex Half-Jacobian Optimization for
N-directional EStimation} (\COH).

Intuitively, the idea relies on that the RIME (Eq. \ref{eq:ME}) has the remarkable property to
behaves like a linear system around the solution. Specifically, from Eq. \ref{eq:Jpq0} and
\ref{eq:Jpq1}, it is easy to check that:

\begin{alignat}{2}
\label{eq:Lin}
\vis=&\frac{1}{2}\left(\JV\big|_{\gwirt}\right)\ \gwirt
\end{alignat}

\noindent while

\begin{alignat}{2}
\label{eq:Lin2}
\vis=&h\left(\vec{g}\right)=\left(\JVg\big|_{\conj{\g}}\right)\ \g\\
\text{and }\vis=&\left(\JVCg\big|_{\g}\right)\ \conj{\g}
\end{alignat}



Both Eq. \ref{eq:Lin} and \ref{eq:Lin2} form linear
systems. Furthermore it is shown in \citet[][]{SmirnovTasse14}
that the principal blocks of $\JV^H\JV$ corresponding to the
$(\vec{g},\conj{\vec{g}})$ and $(\conj{\vec{g}},\vec{g})$ cross
terms have a smaller amplitude than the $(\vec{g},\vec{g})$ and
$(\conj{\vec{g}},\conj{\vec{g}})$ blocks.

\begin{figure*}[t!]
\begin{center}
\includegraphics[width=\textwidth]{\FigDir 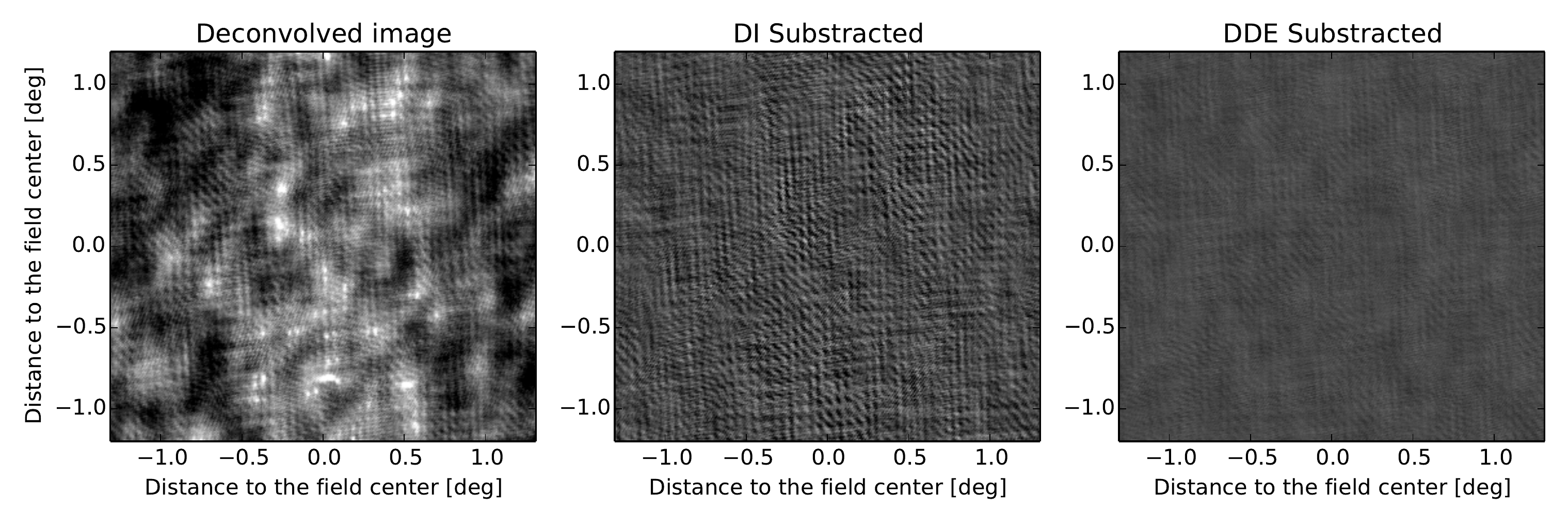}
\caption{\label{fig:resid} This figure shows the comparison between
  the deconvoled image
  (left), the residuals data after simple skymodel subtractions
  (center), and the residuals data after subtracting the
  sky model corrupted by the direction-dependent \COH~estimated solution
  (right). The color scale is the same in all panels. In this
  simulation, \COH~reduces the
residual data level by a factor of $\sim4$.}
\end{center}
\end{figure*}

The structure of $\JVg^H\JVg$ is shown in Fig. \ref{fig:HalfJHJ} for
the dataset described in Sec. \ref{sec:SimpleSimul}. This
matrix is block diagonal, essentially because 
$\partial \conj{g}/\partial g=0$ in Wirtinger's framework. This allows
for dramatic improvement in algorithmic cost, as the $\JV^H\JV$ matrix
inversion cost is $\mathcal{O}\left(n_d^3n_a\right)$ instead of being
$\mathcal{O}\left(n_d^3n_a^3\right)$ corresponding to a net gain of $n^2_a$.

Another interesting property is that, assuming
$\text{diag}\left(\H\right)\approx\H$ and injecting Eq. \ref{eq:Lin2}
into Eq. \ref{eq:LM}, we find:

\def\Fact{\left(\lambda+1\right)^{-1}}
\def\ThisJ{\JVg\big|_{\conj{\widehat{\g_{i}}}}}
\begin{alignat}{2}
\label{eq:HalfLM}
\widehat{\vec{g}_{i+1}}=&\lambda\Fact\widehat{\vec{g}_{i}}+\Fact\H_{\widehat{\vec{g}_{i}}}^{-1}\ThisJ^H\textbf{C}^{-1}\mathbf{v_m}\\
\text{with }\H_{\vec{g}}=&\ThisJ^H\textbf{C}^{-1}\ThisJ
\end{alignat}

\noindent meaning the predict step does not have to be computed along
the iteration.






\section{Tests on simulated data}

In this section, \COH~is tested on simulated data for scalar Jones
matrices only. In Sec. \ref{sec:SimpleSimul}, the
Jones matrices are constant in time. In that case only the convergence
of \COH~can be studied. In Sec. \ref{sec:VarSimul}, the Jones matrices vary in
time.

\begin{figure}
\begin{center}
\includegraphics[width=\columnwidth]{\FigDir 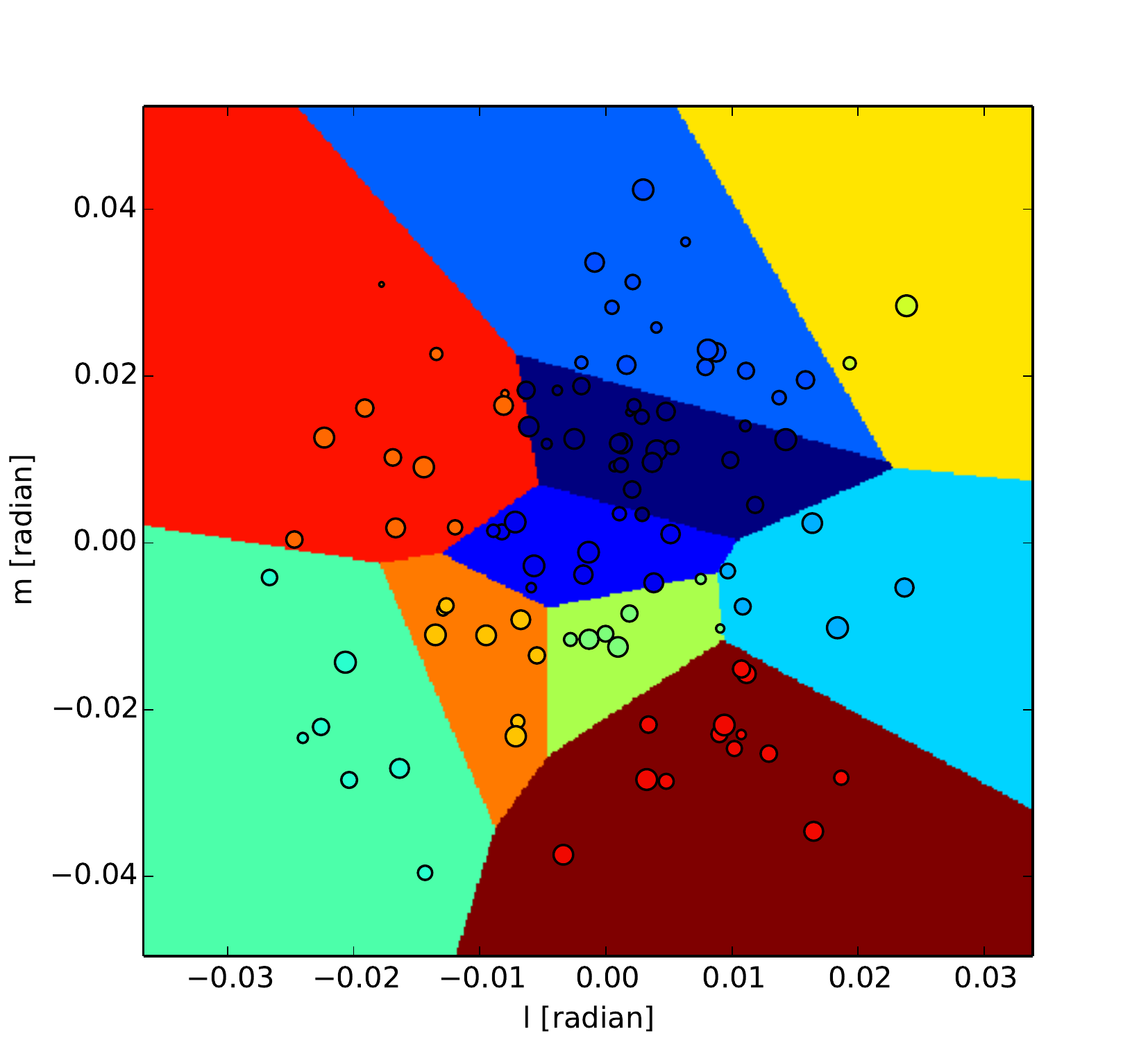}
\caption{\label{fig:tessel} In order to conduct direction-dependent
  calibration, the sources of the sky model are clustered using a Voronoi
  tessellation algorithm.}
\end{center}
\end{figure}

\subsection{Time-constant Jones matrices}
\label{sec:SimpleSimul}

For this test, a visibility dataset is simulated assuming the Low Frequency Array (LOFAR) antenna
layout. The phase center is located at
$(\alpha,\delta)=(14^h11^m20.5^s, +52^{o}12\arcmin10.0\arcsec)$,
the observing frequency is set $50$ MHz, and time
bins are 10 sec wide. To generate the visibilities, we use a sky model containing five
sources, distributed in a cross, and separated by a degree. The gains
applied to the antenna $p$ in direction $d$ are
constant through time, and are taken at random along a normal distribution
$g_{p}\sim\mathcal{N}\left(0,1\right)+i\mathcal{N}\left(0,1\right)$. The
data vector is then built from all baselines, and a $20$ minutes time chunk.

The corresponding matrix $\JVg^H\JVg$ is shown in
Fig. \ref{fig:HalfJHJ}. It is block diagonal, each block having size
$n_d\times n_d$. The calibration solution convergence are shown
in Fig. \ref{fig:Convergence}. It is important to note that the
problems becomes better conditioned as the blocks of $\JVg^H\JVg$
become more diagonal. In that case \COH~converges faster, and this happens (i) when more data are taken into
account in the construction of the data vector or (ii) if the
directions are put further away from each other.

\subsection{Variable gains}
\label{sec:VarSimul}

In order to simulate a more realistic dataset, we use a 100 sources
sky model which flux density is randomly distributed (uniform
distribution). Noise is added to the visibilities at the $1\%$ level
of the total flux. The scalar Jones matrices are simulated assuming an
ionospheric model consisting of a purely scalar, direction-dependent
phase (an infinitesimally thin layer at a height of 100 km). The total
electron content (TEC) values at a set of sample points are generated
using Karhunen-Loeve decomposition \citep[the spatial correlation is
  given by Kolmogorov turbulence, see][]{Tol09}. The sources are
clustered in 10 directions using Voronoi tesselation
(Fig. \ref{fig:tessel}).

Fig. \ref{fig:resid} shows the residuals data computed by subtracting
the model data in the visibility domain, and the model data affected
by DDEs. The residual data standard deviation reduces by a factor
$\sim30$ after \COH~ has been applied.

\section{Conclusion}

This paper has presented a Levenberg-Maquardt based algorithm that
uses the Wirtinger's framework for complex derivative. The Jacobian
harbors a different structure that is sparser than in the classical
case. Based on this, a new optimisation algorithm has been
presented (\COH).

This framework, and its connection with existing algorithms will be
further discussed in \citet{SmirnovTasse14}.


\bibliographystyle{aa}
\bibliography{references}



\end{document}